\documentclass{aa}

\usepackage{graphicx}
\usepackage{natbib}
\usepackage{color}
\usepackage{upgreek}
\usepackage{soul}

\bibpunct{(}{)}{;}{a}{}{,}

\usepackage{txfonts}
\usepackage{mathrsfs}
\usepackage[breaklinks=true]{hyperref}
\begin{document}

\titlerunning{Magnetic blue and yellow stragglers}
\authorrunning{S.~Hubrig et al.}

\title{The incidence of magnetism in blue and yellow straggler stars}

\author{
S.~Hubrig\inst{1}
          \and
          S.~P.~J\"arvinen\inst{1}
          \and
          I.~Ilyin\inst{1}
          \and
          M.~Sch\"oller\inst{2}
                     }
\institute{
Leibniz-Institut f\"ur Astrophysik Potsdam (AIP),
An der Sternwarte~16, 14482~Potsdam, Germany\\
 \email{shubrig@aip.de}
              \and
European Southern Observatory, Karl-Schwarzschild-Str.~2, 85748 Garching, Germany\\            
 }

 
  \abstract
{
  Our understanding of the generation of magnetic fields in intermediate-mass and massive OBA stars
  remains limited. Some theories have proposed that their magnetic fields could be a result of strong binary
  interactions, including stellar mergers. Blue straggler stars, which lie well beyond the main sequence
  turn-off point on the colour-magnitude diagram of stellar clusters, are widely theorised to be merger
  products or interacting binaries and therefore can be considered excellent test targets to get insights into the origin
  of magnetic fields in stars with radiative envelopes.
  }
   {
     We search for the presence of magnetic fields in a sample of blue and yellow straggler stars
     listed in the Gaia DR2-based catalogue of blue straggler stars in open clusters.
}
{
  We measured the mean longitudinal magnetic field
  from high-resolution HARPS\-pol spectra
of five blue straggler and three yellow straggler stars using the least-squares deconvolution technique.
}
{We present the first observational evidence that blue straggler and  yellow straggler stars possess magnetic fields of the order of
a hundred to a few hundred Gauss. The targets in our sample belong to open clusters of very different ages
and metallicities, but we do not detect any relationship  between the presence or  strength of the detected
magnetic field and the cluster characteristics.
For the first time, using high-resolution spectropolarimetric observations,
  a definite detection of a magnetic field is achieved in a Be-shell star, HD\,61954.
The two yellow straggler stars, HD\,62329 and HD\,65032, appear to be members in binary systems, whereas the blue straggler star HD\,62775
  is possibly a triple system.
HD\,62329 and HD\,65032 exhibit in their spectra weak
  \ion{Nd}{iii}~6145 lines, which are usually prominent in magnetic Ap and Bp stars. 
 Our observations provide crucial information necessary for testing predictions of existing theories and
place strong constraints on the origin of magnetic fields in stars with radiative envelopes.
}
   {}

   \maketitle
%

   \section{Introduction}
   \label{sect:intro}

   The existence of magnetic fields in massive and intermediate-mass OBA stars, with an incidence of
   roughly 10--15\%, is no longer in question. However, the origin of these fields is still unclear.
It has been argued in the past that magnetic fields in stars with radiative envelopes could be fossil
relics of fields that were present in the interstellar medium from which these stars formed (e.g.\ \citealt{Moss2003}).
Alternatively, magnetic fields may be generated by strong binary interactions, i.e.\ in the course
of mass transfer, during common envelope evolution, or during a merger of two lower-mass stars or protostars
(e.g.\ \citealt{Tout2008,Ferrario2009,Tutukov2010}).
Mass transfer or stellar mergers may rejuvenate the mass-gaining star, while
the induced differential rotation is thought to be the key ingredient for the generation of a magnetic field
(e.g.\ \citealt{Wickramasinghe2014}). Three-dimensional magnetohydrodynamical simulations  by \citet{Schneider2019}
demonstrated that merger products may exhibit strong magnetic fields and rapid rotation immediately
following the merger.
Although a few studies have already reported on the presence of magnetic fields in interacting binaries or merger products (e.g.\
\citealt{Hubrig2022,Hubrig2023,Frost2024}), 
dedicated spectropolarimetric observations are necessary to secure trustworthy statistics on the occurrence of
magnetic fields and the distribution of field strengths in interacting binaries.

One of the observable manifestations of the presence of coalesced stars or 
rejuvenated companions in binary systems due to mass transfer is the existence of
blue straggler stars (BSSs) in open clusters (OCs). These stars, due to their locations in the colour-magnitude
diagram (CMD) beyond the turn-off point (e.g.\ \citealt{Rain2021}) appear more luminous, hotter, and
therefore younger than their coeval counterparts. \citet{McCrea1964} proposed that mass transfer from a
binary companion can lead to rejuvenation of the mass gainer and the formation of a BSS. 
More recent studies have shown that tight binary stars that merged as a consequence of Roche-lobe overflow of both
of their components during the hydrogen-burning evolution form
rejuvenated stars that could be brighter and bluer than the turn-off stars in star clusters
(e.g.\ \citealt{Wang2020,Wang2024}). In hierarchical triple systems, the Kozai-Lidov mechanism can cause the inner binary to
become more compact (e.g.\ \citealt{Naoz2014}), leading to mergers within the inner binary system and
contributing significantly to BSS formation in OCs.
Obviously, given the results of recent theoretical simulations demonstrating that binary interaction products 
exhibit measurable magnetic fields and the fact that the primary mechanisms for BSS formation include
binary mass transfer, binary mergers, and stellar collisions, a systematic survey of magnetic fields
in BSSs is urgently needed.

Indeed, the question of the presence of magnetic fields in BSSs is not answered yet.
\citet{Pendl1976} found from spectroscopy that at least one third of the 14 studied BSSs had spectra similar
to magnetic Ap and
Bp stars. \citet{Maitzen1981} conducted photometric studies of BSSs and found that
around 10\% of BSSs show chemical peculiarities inherent to Ap stars. Other studies that arrived at this conclusion
include \citet{Abt1985} and \citet{Paunzen2014}.  \citet{Mathys1988}
failed to detect magnetic fields in four BSSs in one of the oldest Galactic clusters, M67, using measurements with
high uncertainties – of the order of several hundred Gauss.
\citet{Hubrig2008} studied the magnetic field of the SB1 hot peculiar B0.2V system $\theta$~Carinae (=HD\,93030),
suggested to be a BSS located $\sim$2\,mag above the turn-off of the young OC IC\,2602, but
the results were
inconclusive, with only a few measurements at a significance level of $3\sigma$. Previous studies of this binary
with one of the shortest orbital periods ($P=2.2$\,d) known among massive stars
\citep{Lloyd1995} reported the presence of spectral
peculiarities such as an enhancement of nitrogen and definite line-intensity variations
\citep{Walborn1979} indicative of previous mass transfer. On the other hand, due to its brightness rendering astrometry
unreliable, this system was not listed in the recent
study of \citet{Rain2021}, who used {\it Gaia} DR2 data to establish the membership of BSSs in 408 Galactic OCs.

With respect to more recent studies, \citet{Nieva2014} measured  the stellar parameters
of the magnetic massive B0.2\,V star $\tau$~Sco
belonging to the Upper Sco association and, based on its position in the HRD,
concluded that this star is a BSS and could possibly originate from a stellar merger.
Also the most recent kinematical study of the young
OC NGC\,2516 by \citet{Kharchenko2022} based on high-precision Gaia EDR3 data
confirmed membership status (and likely BSS
status) of the magnetic chemically peculiar Bp star HD\,65987. Since the Ohmic decay timescale is
comparable to the stellar lifetime, the generated magnetic fields are expected to be long-lived.

In light of these results, it is obvious that spectropolarimetric observations of BSSs are important 
to test whether the presence of a magnetic field correlates with the blue straggler phenomenon.
Therefore, as a pilot project, we obtained twelve spectropolarimetric
HARPS\-pol observations to search for magnetic fields in six BSSs and two yellow straggler stars (YSSs)
identified in various clusters and presented in the catalogue of
\citet{Rain2021}. This catalogue is
based on photometry, proper motions, and parallaxes extracted from Gaia DR2. The two YSSs, which are
believed to be stragglers in a more advanced evolutionary stage, have been observed to test the evolutionary effect
on the strength of the magnetic field. To our knowledge, no search for magnetic fields has ever been 
carried out for any YSS in the past. Obviously, observations of YSSs are important to understand the evolution
of magnetic fields across the CMD.

 The paper is laid out as follows.
 In Sect.~\ref{sect:sam} we describe the targets selected for observations.
  In Sect.~\ref{sect:mfield} we present our HARPS\-pol spectropolarimetric observations, their reduction,
 and the results of the magnetic field measurements.
 The occurrence of magnetic fields and the distribution of their strength in BSSs and YSSs
 in the context of recent theoretical considerations of the blue and yellow straggler phenomenon are discussed in
 Sect.~\ref{sect:disc}.


\section{The sample} 
  \label{sect:sam}

We searched for the presence of magnetic fields in BSSs
and YSSs identified in the Gaia-based catalogue of BSSs in open
clusters by \citet{Rain2021}. Before Gaia data became available, the most comprehensive
list of BSSs had been produced by \citet{Ahumada2007}. As we wish to explore the existence of magnetic fields in
massive and intermediate-mass OBA stars, we considered BSSs and YSSs in rather young clusters with ages
from a few Myr to about 700\,Myr.
According to \citet{Jadhav2021}, BSS-like systems are more prevalent in more massive, denser, and older clusters.  The authors also report on the large percentage of BSSs known to be in binaries.
A detection of BSSs requires a clear identification of the main-sequence turn-off of the clusters,
which is much easier for older clusters. \citet{Ahumada2007} demonstrated the difficulty of identifying BSSs in young clusters
without using a corresponding isochrone. On the other hand, the choice of a set of isochrones implies the adoption of
specific stellar models.

Our sample of spectropolarimetrically studied stars
consists of one BSS (HD\,61954) and two YSSs (HD\,62000 and HD\,62329) in the cluster NGC\,2437, one BSS (HD\,62775) in the
cluster NGC\,2447, one BSS (HD\,65032) in the cluster Trumpler\,9, two BSSs (HD\,87222 and HD\,87266) in the cluster
NGC\,3114, and one BSS (HD\,101545) in the cluster IC\,2944.
As is discussed in Sect.~\ref{sect:mfield.trumpler9}, HD\,65032 is listed as a BSS in the
catalogue of \citet{Rain2021}, but has a Gaia $G$ magnitude and colour $BP-RP$ consistent with YSS status \citep{Gaia2022}.
HD\,101545 was identified as the only BSS in the very young
OC IC\,2944 by \citet{Ahumada2007}. However, this cluster was not included
in the catalogue of \citet{Rain2021}, probably because of limitations set by photometric calibration errors for bright targets.

The targets were selected from clusters with ages between
 70 and 700\,Myr. 172 clusters in the catalogue of \citet{Rain2021} have ages in this range,
of which 108 are observable from
La~Silla, i.e.\ with declinations below $+20^{\circ}$.
From the same catalogue, we find a total of 24 BSSs with $m_{\rm V}$ 
smaller or equal to 10, of which 17 are below $+20^{\circ}$, of which 13
are from clusters with an age between 70 and 700\,Myr.
For the YSS, these numbers are 8, 7, and 6.
Due to the younger age of clusters containing OBA stars, only very few targets have been classified as YSSs
by \citet{Rain2021}.
There is no cluster younger than 70\,Myr holding a YSS and
there are only two clusters younger than 70\,Myr holding at least one BSS:
NGC\,7790 with one and IC\,361 with two.
All three sources are faint with $m_{\rm V}$ > 10.
To test possible correlations between the occurrence of
magnetic fields in BSSs and YSSs and cluster properties, our targets are selected from OCs of different ages
and metallicities. In Table~\ref{tab:clusters} we present for each cluster the corresponding age and metallicity together with
the literature source.

\begin{table*}
\centering
\caption{
  Parameters of the clusters harbouring our BSSs and YSSs. 
}
\label{tab:clusters}
\begin{tabular}{lr@{$\pm$}lr@{$\pm$}ll}
\hline
\hline\noalign{\smallskip}
\multicolumn{1}{c}{Cluster}   &
\multicolumn{2}{c}{log(age)}   &
\multicolumn{2}{c}{[Fe/H]}    &
\multicolumn{1}{c}{Reference} \\
\multicolumn{1}{c}{}      &
\multicolumn{2}{c}{[yr]}      &
\multicolumn{2}{c}{[dex]} &
\multicolumn{1}{c}{}      \\
\hline\noalign{\smallskip}
NGC~2437        & 8.709 & 0.049               & 0.011    & 0.071          & \citet{Dias2021} \\
NGC~2447        & 8.825 & 0.029               & $-$0.051 & 0.010          & \citet{Dias2021} \\
NGC~3114        & 8.358 & 0.055               & 0.096    & 0.039          & \citet{Dias2021} \\
Trumpler~9      & 7.869 & 0.263               & $-$0.083 & 0.053          & \citet{Dias2021} \\
IC~2944         & \multicolumn{2}{c}{$\sim$7} & \multicolumn{2}{c}{solar} & \citet{Baume2014} \\
\hline
\end{tabular}
\tablefoot{
In the first column we give the name of the cluster. The ages and metallicities are
  listed in columns~2 and 3, followed by the literature source.
}
\end{table*}

\section{Observations and  magnetic field measurements}
  \label{sect:mfield}

\begin{table*}
\caption{
Logbook of the observations and the results of the magnetic field
measurements for all targets in our sample. 
}
\label{tab:obsall}
\centering
\begin{tabular}{rcccrccc r@{$\pm$}l c}
\hline\hline \noalign{\smallskip}
\multicolumn{1}{c}{HD} &
\multicolumn{1}{c}{Spectral} &
\multicolumn{1}{c}{$m_{\rm V}$} &
\multicolumn{1}{c}{MJD} &
\multicolumn{1}{c}{$S/N$} &
\multicolumn{1}{c}{Line} &
\multicolumn{1}{c}{FAP}&
\multicolumn{1}{c}{Det.} &
\multicolumn{2}{c}{$\left< B \right>_{\rm z}$} &
\multicolumn{1}{c}{Remark} \\
\multicolumn{1}{c}{number} &
\multicolumn{1}{c}{type} &
\multicolumn{1}{c}{} &
\multicolumn{1}{c}{} &
\multicolumn{1}{c}{} &
\multicolumn{1}{c}{mask} &
\multicolumn{1}{c}{} &
\multicolumn{1}{c}{flag} &
\multicolumn{2}{c}{[G]} &
\multicolumn{1}{c}{} \\
\noalign{\smallskip}\hline \noalign{\smallskip}
61954  & B9$^\ast$ &9.44  & 60\,311.14 & 127 & Si        & $7.4\times10^{-5}$  & MD & $-$420  & 95   & BSS, NGC\,2437\\
       &           &      & 60\,315.06 &  98 & HeCSi   & $2.8\times10^{-6}$  & DD & 156     & 46   & \\
62000  & A1II/III  &9.23  & 60\,311.27 & 156 & Fe        & $<10^{-10}$       & DD & $-$4    & 29   & YSS, NGC\,2437\\
       &           &     &            &     & Ti        & $1.4\times10^{-5}$  & MD & $-$111   & 43  & \\
62329  & A2$^\ast$ & 9.24 & 60\,312.24 & 121 & Fe        & $9.3\times10^{-5}$ & MD & 114    & 87  & YSS, NGC\,2437\\
       &            &       & 60\,314.09 &  96 & Fe        & $1.5\times10^{-6}$  & DD &$-$582  & 86 &   \\
62775  & A2III/IV & 8.12  & 60\,312.06   & 133 & Fe        & $6.4\times10^{-5}$  & MD & 631  & 110 &  BSS, NGC\,2447 \\
         &          &      & 60\,314.12  & 182  & FeTi     & $1.1\times10^{-7}$  & DD & \multicolumn{2}{c}{}&   \\
65032  & A2/3III    & 8.35  & 60\,312.11 & 126 & Fe        & $0.5\times10^{-5}$  & MD & \multicolumn{2}{c}{}& likely YSS, Trumpler\,9\\
       &            &   & 60\,313.09 & 132 & SiCr     & $7.8\times10^{-8}$  & DD & \multicolumn{2}{c}{}  & \\
87222  & B3IV     & 8.63 & 60\,313.16 & 137 & HeC   & $8.1\times10^{-4}$       & MD & $-$133  & 7  & BSS, NGC\,3114 \\
87266  & B3III/V    &8.23  & 60\,315.19 & 164 & CNNeSi      & $1.0\times10^{-5}$  & DD & 110   & 23   & BSS, NGC\,3114\\
101545\,A & O9.5Ib &6.38 & 60\,311.32 & 194 & HeNOSi & $2.1\times10^{-5}$  & MD & 69      & 17   & BSS, IC\,2944\\
                          &   &   &          &  & He                        & $4.3\times10^{-6}$  & DD & 82   & 16    & \\
\hline
\end{tabular}
\tablefoot{
  The first column gives the HD number of the star followed by the spectral type taken either from SIMBAD or from
 \citet{Cannon1993} (marked by $^\ast$)
and the visual magnitude. The fourth column presents
the MJD values at the middle of the exposure, while in the fifth column we
show the signal-to-noise ratio measured in the Stokes~$I$ spectra in the
spectral region around 5000\,\AA. The applied line masks, the FAP values, the
detection flag -- where DD means definite detection, MD marginal detection,
and ND no detection -- the measured LSD mean longitudinal magnetic field
strength, and a remark on cluster membership are presented in columns 6–10.
}
\end{table*}

Our spectropolarimetric observations were carried out using HARPSpol attached to ESO’s 3.6m telescope in 2024
from January~1 to January~6.
With HARPS\-pol, we have access to measurements of the mean longitudinal magnetic field $\langle B_{\rm z}\rangle$,
which is the line-of-sight component of the magnetic field, weighted with the line
intensity and averaged over the visible hemisphere. The longitudinal magnetic field is strongly dependent on the
viewing angle between the field orientation and the observer and is modulated as the star rotates.
HARPS\-pol has a resolving power of about 115\,000 and a wavelength coverage from
3780 to 6910\,\AA{}, with a small gap between 5259 and 5337\,\AA{}.
The data were reduced on La~Silla using the HARPS\-pol data reduction pipeline.
The normalisation of the spectra to the continuum level is described in detail in \citet{Hubrig2013}.
Due to a misalignment of the polarimeter, the achieved signal-to-noise ratio ($S/N$) was somewhat lower than intended.
This problem went unnoticed over a couple of years, because this misalignment did not affect the polarimetric analysis itself
and only the overall transmission was decreased. The ESO staff successfully fixed the position of the polarimeter in March 2025.

As we did in our previous studies using HARPS\-pol data (e.g.\ \citealt{Hubrig2018,Jarvinen2020}),
we employed the least-squares deconvolution (LSD) technique following the description given by \citet{Donati1997}
in order to increase the accuracy of the mean longitudinal magnetic field determination.
The parameters of the lines used to calculate the LSD profiles were taken
from the Vienna Atomic Line Database \citep[VALD3;][]{Kupka2011}.
Only lines that appear to be unblended or minimally
blended in the Stokes~$I$ spectra were included in the line mask.
The resulting profiles were scaled according to the line strength and
the sensitivity to the magnetic field. Because early-type stars are frequently members of binary or multiple systems,
for the treatment of their composite spectra a special procedure 
involving different line masks populated for each element separately has been applied. This procedure is similar to that
described by \citet{Hubrig2023} in their study of magnetic fields in massive binary and multiple systems. 

Such a strategy is also important for stars hosting magnetic fields.
  Usually, abundance anomalies in magnetic stars with radiative envelopes are associated with
the presence of surface chemical patches, which show a correlation with
magnetic field topologies (e.g.\ \citealt{Rice1997,Hubrig2012,Hubrig2017a,Hubrig2017b}).
As their visibility is changing during the stellar rotation,
different line masks populated for each element separately should be tested.

The presence of a magnetic field in the LSD profile was evaluated according to \citet{Donati1992},
who defined a Zeeman profile with a false alarm probability (FAP) $\leq 10^{-5}$ as a definite detection, $10^{-5} <$ FAP $\leq 10^{-3}$ as a marginal detection, and FAP $> 10^{-3}$ as a non-detection.
The FAP values and the longitudinal magnetic field strengths were calculated in the velocity ranges corresponding to the full
width of the underlined Stokes $I$ profiles. Furthermore, to check the reliability of our results, 
we also calculated FAP values for the $N$ spectra. All our measurements yield non-detections with the smallest FAP value of
0.15 achieved for HD\,61954.
To avoid the limitations to the accuracy of the magnetic field characterisation due to the observed longitudinal
magnetic field changing with rotational and/or orbital phase, a few targets have been observed twice.
In this way we sampled different phases and hoped to gain a more refined understanding of the field geometry.
The summary of our measurements is presented in Table~\ref{tab:obsall}.
In the following subsections, we discuss the spectral appearance of each target, including  previously
known properties and the results of our measurements.

\subsection{NGC\,2437}

\paragraph*{HD\,61954 (=BD$-$14\,2102):}

\begin{figure*}
    \centering
    \includegraphics[width=0.75\textwidth]{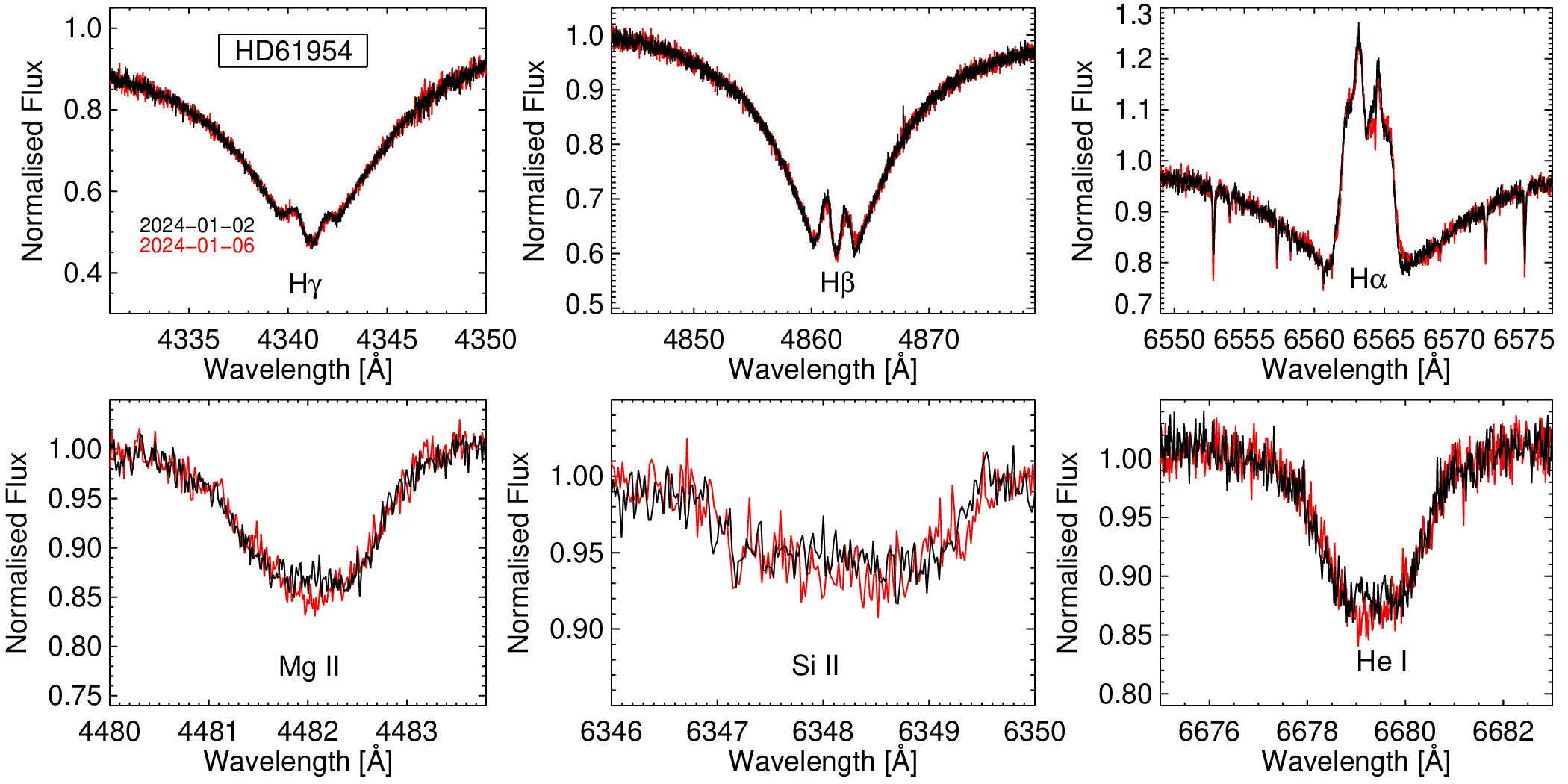}
    \caption{
Line profiles of various lines observed in the HARPS\-pol spectra of HD\,61954 obtained on two different nights.
   }
    \label{fig:reg61954}
\end{figure*}

\begin{figure}
    \centering
    \includegraphics[width=0.237\textwidth]{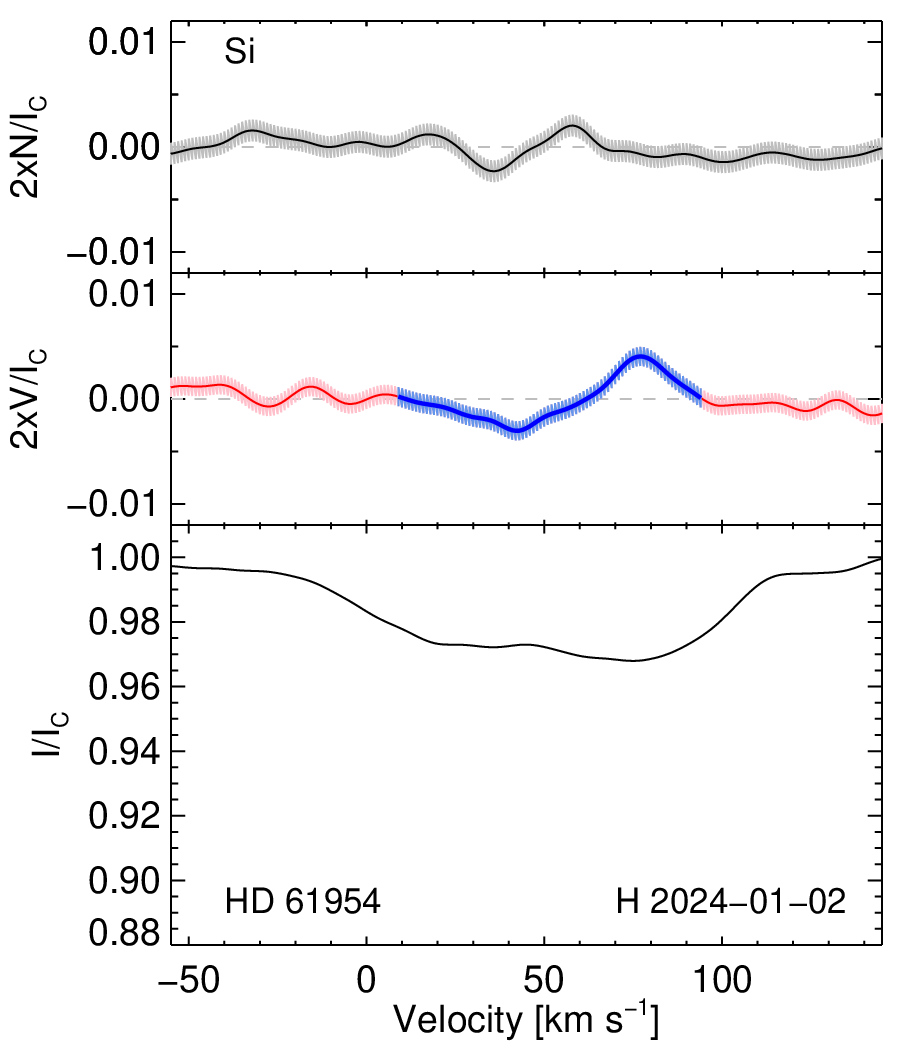}
    \includegraphics[width=0.237\textwidth]{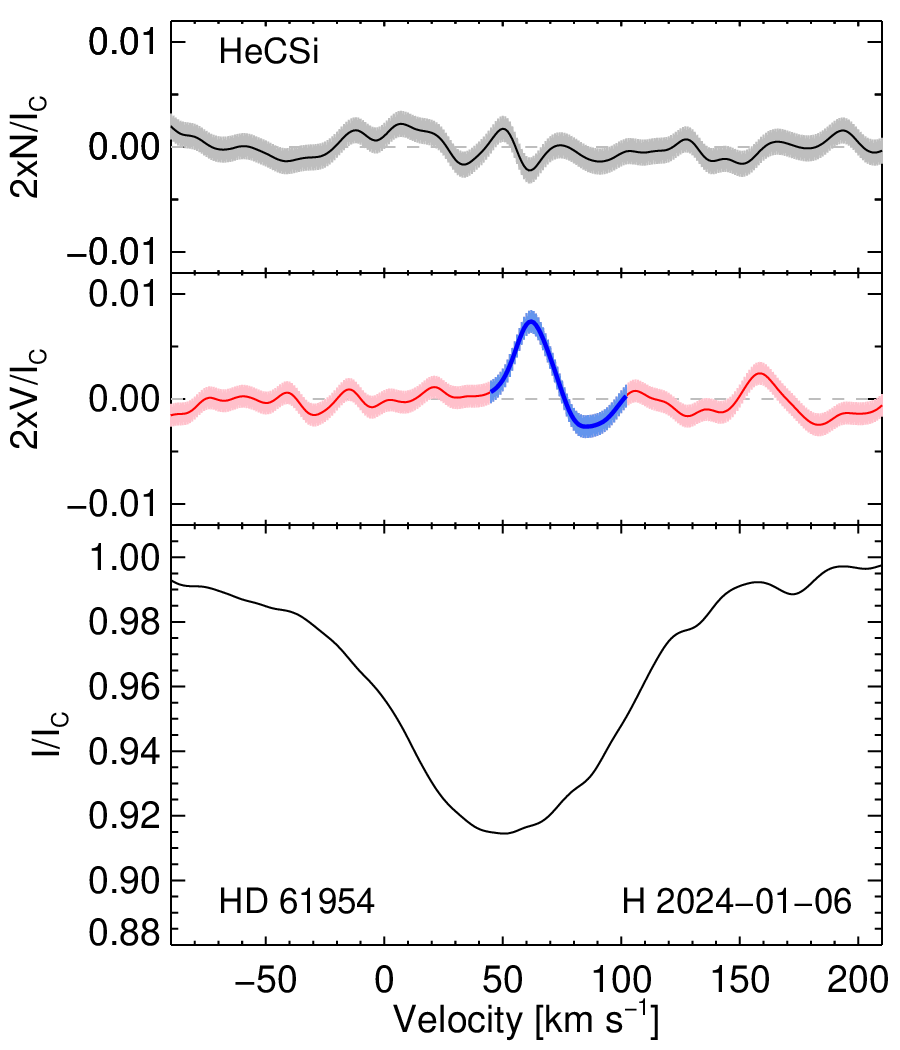}
    \caption{LSD analysis results, Stokes~$I$, $V$, and diagnostic null $N$ spectra, obtained for
      different line masks using HARPS\-pol observations of HD\,61954 on two different nights in January 2024. The
        identified Zeeman signatures are presented in a blue colour.
}
    \label{fig:IVN61954}
\end{figure}

With only 14 entries in the SIMBAD database, not much is known about this target with a spectral type of B9 listed in \citet{Cannon1993}.
According to \citet{Rain2021}, it is a BSS in the OC NGC\,2437 with a probability of 0.8.
In Fig.~\ref{fig:reg61954} we present line profiles of various lines observed in the HARPS\-pol spectra obtained on two different nights.
  The appearance of emission in the hydrogen lines suggests that this target belongs to the class of Be-shell stars \citep{Rivinius2006}.
  Although we do not detect a radial velocity change between the two different observing epochs, small changes in the line profile shapes of 
\ion{Mg}{ii}~4481 and \ion{He}{i}~6678 still may indicate binarity or a spotted surface. Furthermore, with just two
observations of this target, we cannot exclude the presence of pulsational variability as Be stars are known to exhibit
non-radial pulsations causing variations in light and spectral line profiles. The reason for the detection of small changes in line profiles
can be determined only after
spectropolarimetric monitoring is carried out over the rotation period, which is currently unknown.

Our LSD analysis of observations acquired on two different nights is presented in Table~\ref{tab:obsall}
and Fig.~\ref{fig:IVN61954}.
The LSD analysis of the first  spectropolarimetric
observation on January~2 2024 using a mask containing Si lines shows a marginal detection of a mean longitudinal magnetic field,
$\left< B_{\rm z} \right>=-420\pm95$\,G, with ${\rm FAP}=7.4\times10^{-5}$.
For the second observation on January 6, using He, C, and Si lines
we obtained a definite detection of $\left< B_{\rm z} \right>=156\pm46$\,G  with ${\rm FAP}=2.8\times10^{-6}$.
Importantly, for the observations on the second epoch we observe in Fig.~\ref{fig:IVN61954} that the detected Zeeman
signature does not extend over the full LSD Stokes~$I$ profile. This is explained by the fact that magnetic A- and B-type stars
are characterised by an inhomogeneous chemical abundance distribution that is non-uniform and non-symmetric with
respect to the rotation axis but that shows some symmetry between the topology of the magnetic field and
the chemical spot distribution. Due to the presence of a rotationally modulated appearance of chemical spots in the observed spectra,
the corresponding Zeeman signatures usually do not extend over the full LSD Stokes~$I$ profile. 
The diagnostic null ($N$)
spectrum for the first observation shows a distinct feature, which could be caused by pulsations,
but the FAP value calculated for this spectrum indicates non-detection.

\paragraph*{HD\,62000 (=ALS\,661):}

\begin{figure}
    \centering
    \includegraphics[width=0.7\columnwidth]{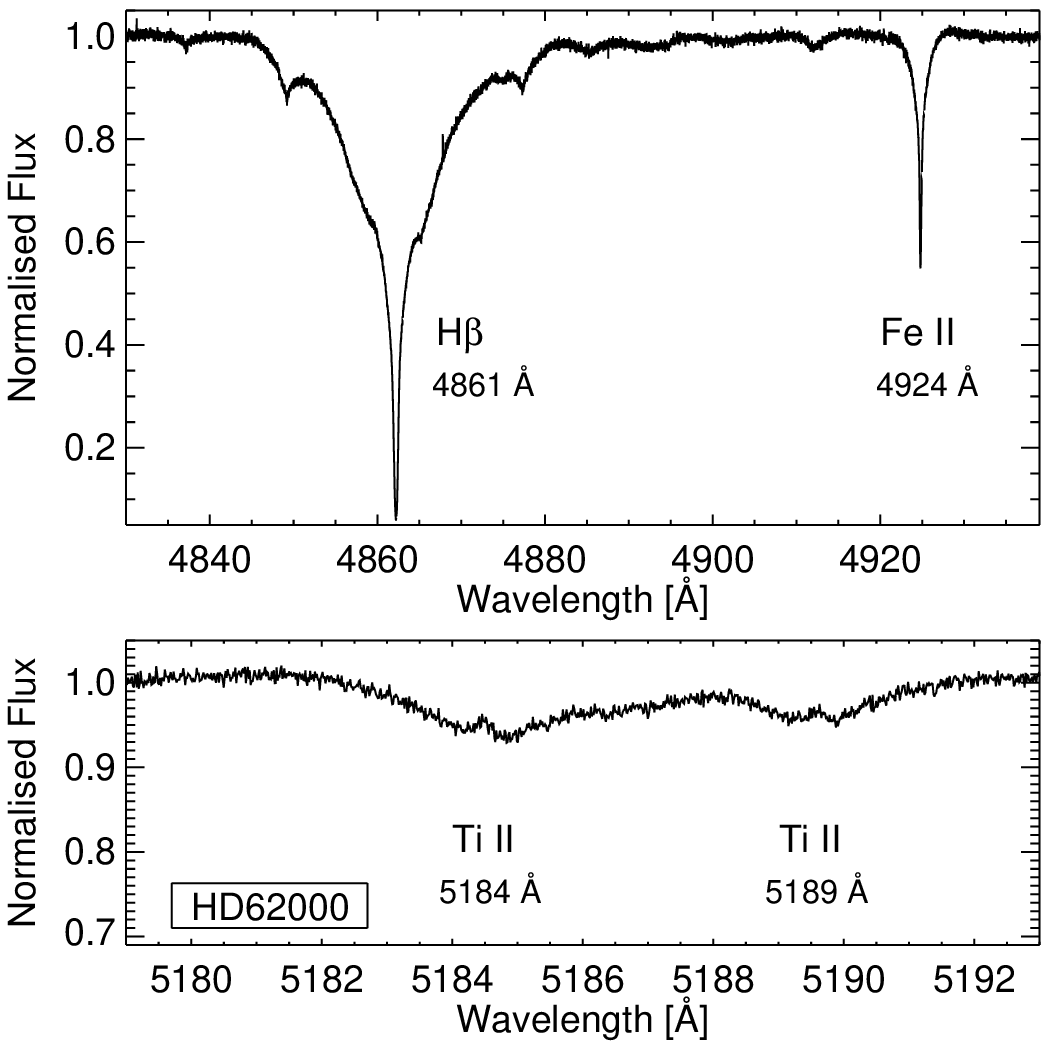}
    \caption{Line profiles belonging to different elements with very unusual shapes observed in the
      HARPS\-pol spectrum of HD\,62000 recorded on January~2 2024.
}
    \label{fig:regH62000}
\end{figure}

\begin{figure}
    \centering
    \includegraphics[width=0.237\textwidth]{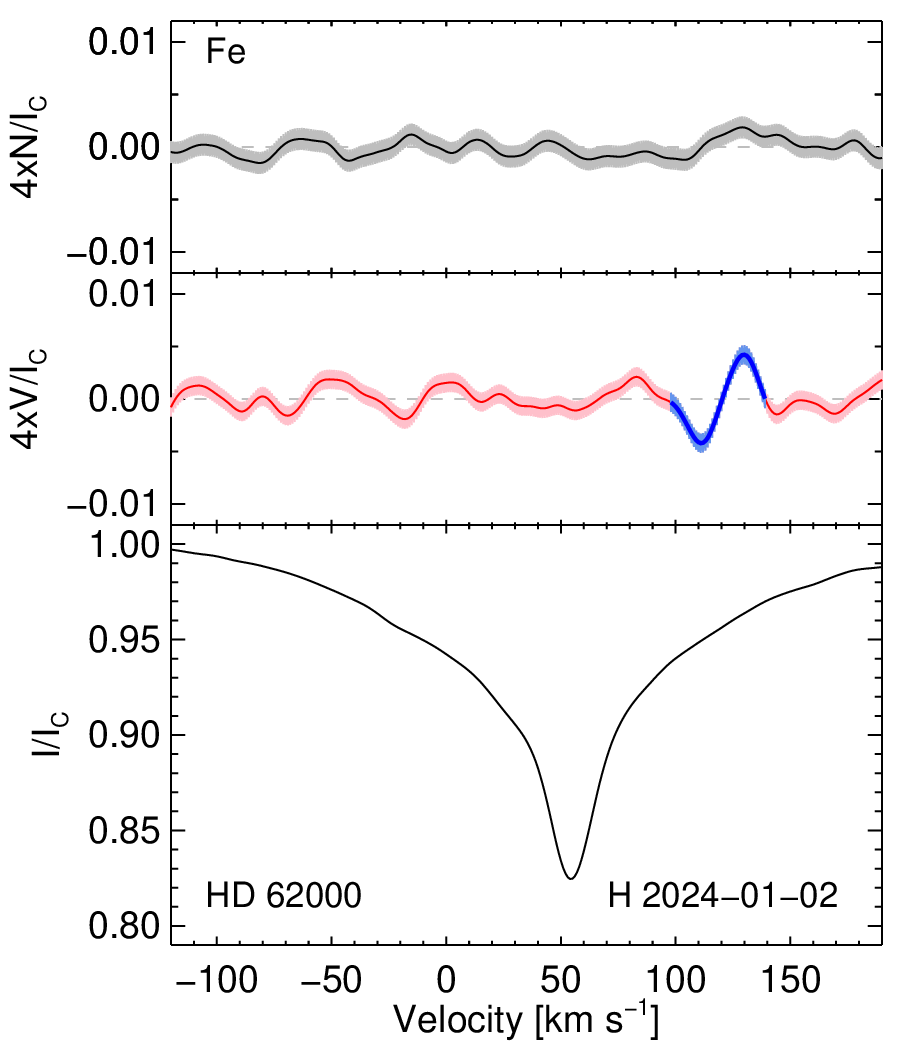}
    \includegraphics[width=0.237\textwidth]{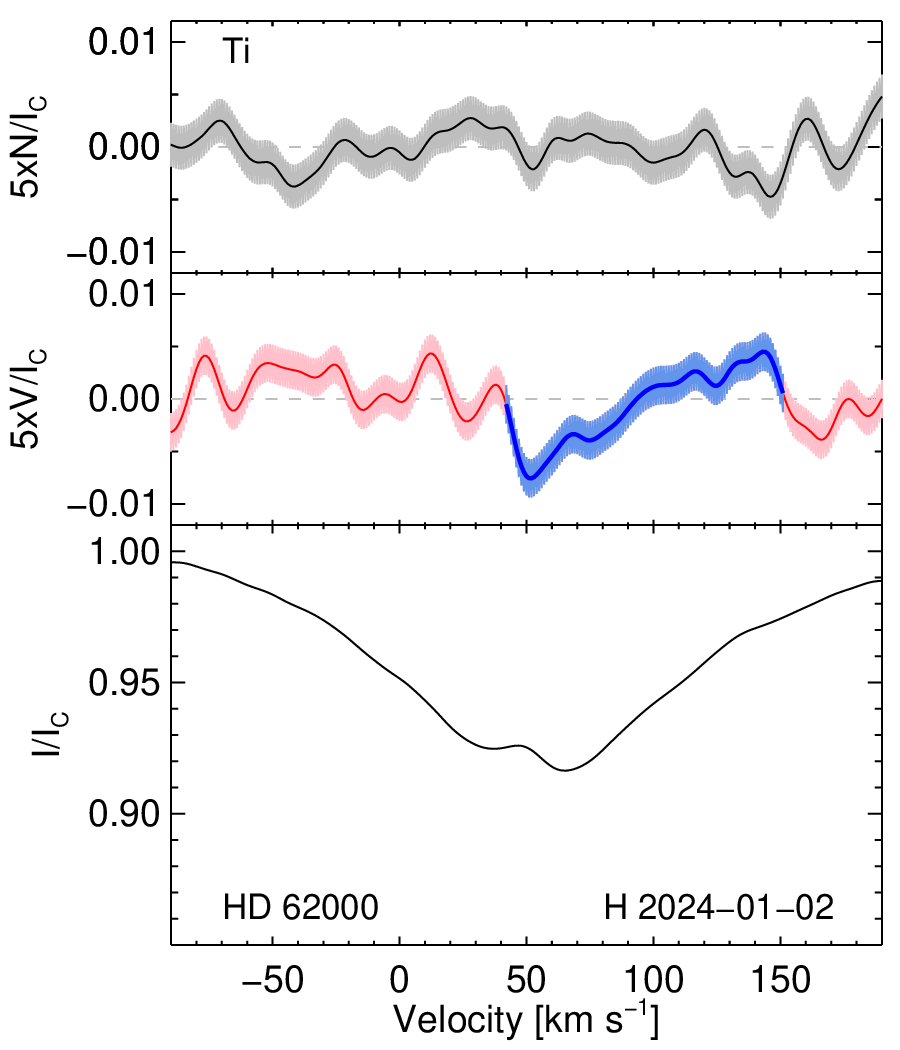}
    \caption{As Figure~\ref{fig:IVN61954}, but for HD\,62000.}
    \label{fig:IVNH62000}
\end{figure}

This target with spectral type A1\,II/III is listed in the catalogue
of \citet{Rain2021} as a YSS in NGC\,2437 at a membership probability of 1.0.
Our inspection of the 
HARPS\-pol spectrum recorded on January~2 2024 suggests a shell phenomenon revealing a very unusual appearance of metal line profiles:
hydrogen and Fe lines exhibit shell-like line profiles sometimes observed in $\lambda$~Boo stars and the Ti line profiles appear broad and split.
The corresponding line profiles are displayed in Fig.~\ref{fig:regH62000}.
Given the very different appearance of the line profiles belonging to different elements, we used in our LSD analysis
two different masks, one containing exclusively Fe lines and another one with Ti lines.
The results of our analysis are presented in Table~\ref{tab:obsall} and in Fig.~\ref{fig:IVNH62000}.
The LSD analysis carried out using Fe lines shows a definite detection $\left< B_{\rm z} \right>=-4\pm29$\,G with
${\rm FAP}<10^{-10}$. For the Ti mask we obtained $\left< B_{\rm z} \right>=-111\pm43$\,G with ${\rm FAP}=1.4\times10^{-5}$.

\paragraph*{HD\,62329 (=BD$-$14 2166):}

\begin{figure}
    \centering
    \includegraphics[width=0.237\textwidth]{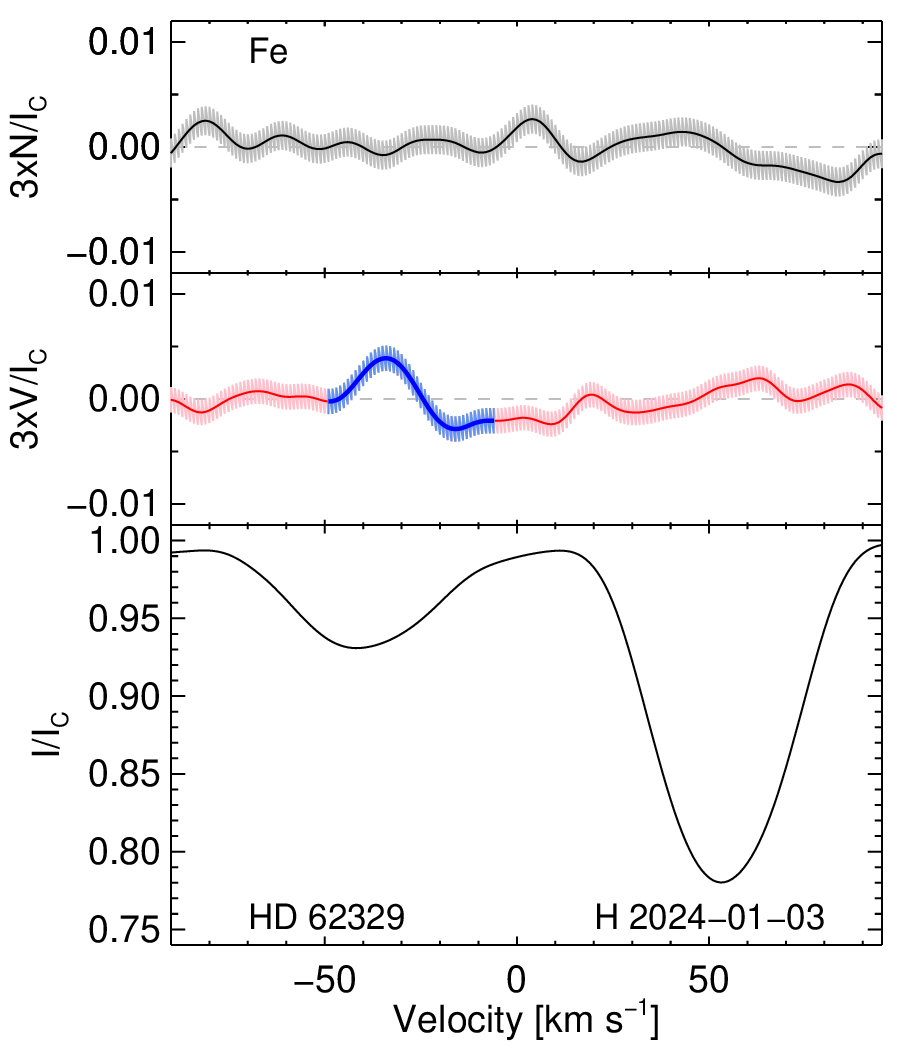}
    \includegraphics[width=0.237\textwidth]{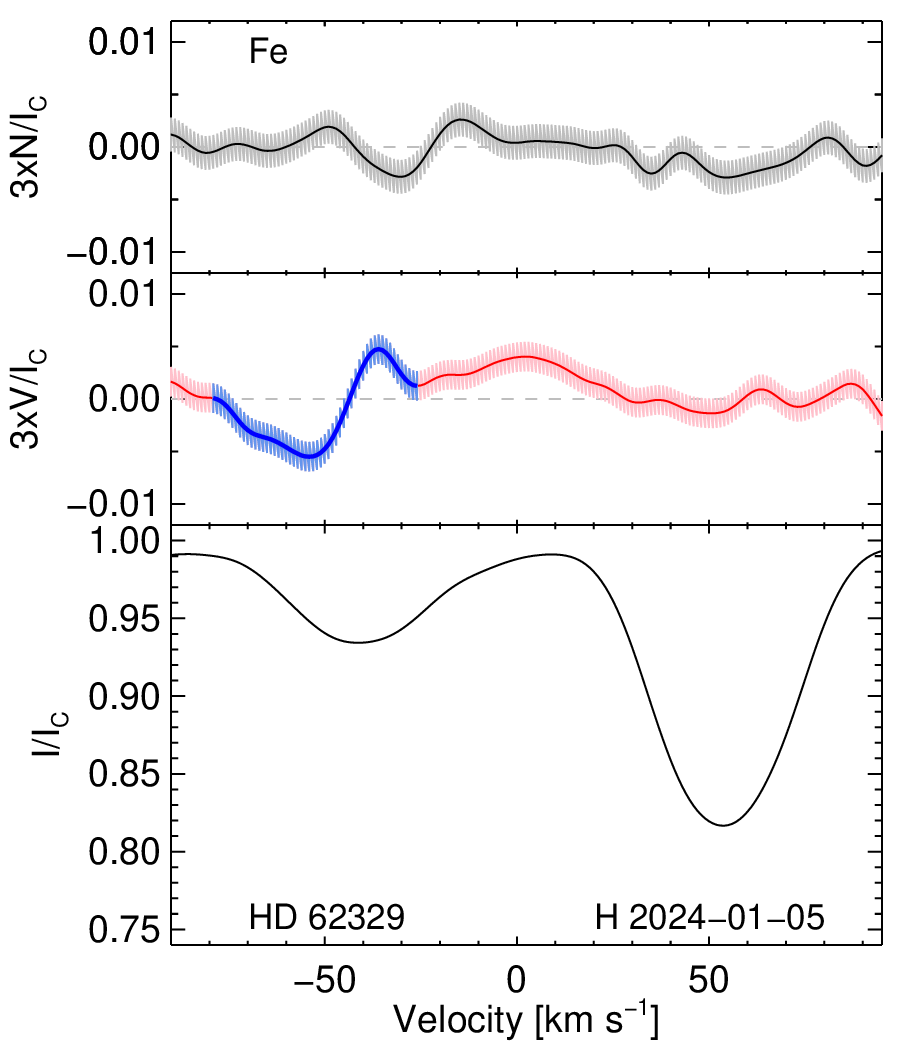}
    \caption{As Figure~\ref{fig:IVN61954}, but for HD\,62329.
}
    \label{fig:IVNHD62329}
\end{figure}

With only seven entries in the SIMBAD database, not much is known about this target with the spectral type A2 listed in \citet{Cannon1993}
It is reported to be a YSS in NGC\,2437 with a membership probability of 0.5 in the catalogue
of \citet{Rain2021}.
For both observations, acquired on January~3 and 5 2024, we used a line mask containing Fe lines.
As is reported in Table~\ref{tab:obsall}, we achieve for the first observation a marginal detection of
$\left< B_{\rm z} \right>=114\pm87$\,G with ${\rm FAP}=9.3\times10^{-5}$ for the night of January~3 and a definite detection
$\left< B_{\rm z} \right>=-582\pm86$\,G with ${\rm FAP}=1.5\times10^{-6}$ for the night of January~5.
The LSD Stokes~$I$ profiles clearly show that HD\,62329 is a double-lined spectroscopic binary with a magnetic component.
The LSD Stokes~$I$, $V$, and diagnostic null $N$ profiles obtained using HARPS\-pol observations obtained on January~3 and 5 2024 are presented in Fig.~\ref{fig:IVNHD62329}.

\subsection{NGC\,2447}

\paragraph*{HD\,62775 (=CD$-$23\,6092):}

  \begin{figure*}
    \centering
    \includegraphics[width=0.245\textwidth]{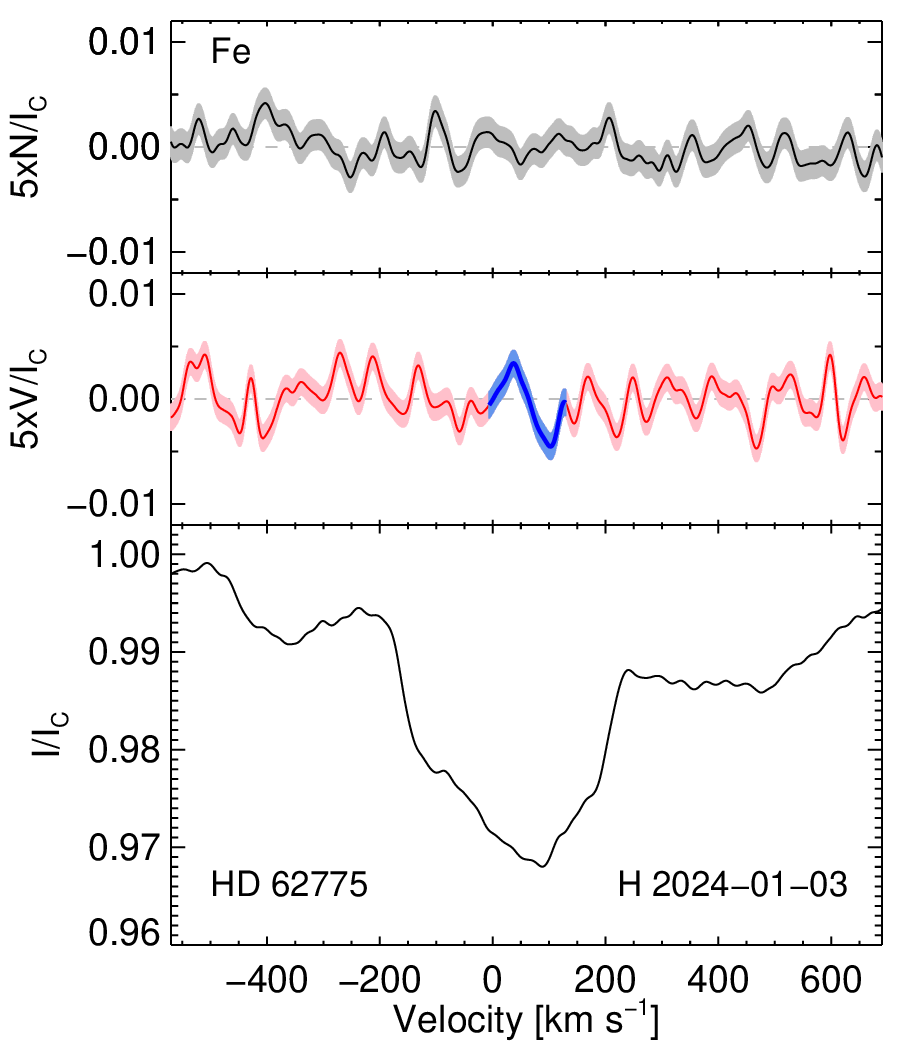}
    \includegraphics[width=0.245\textwidth]{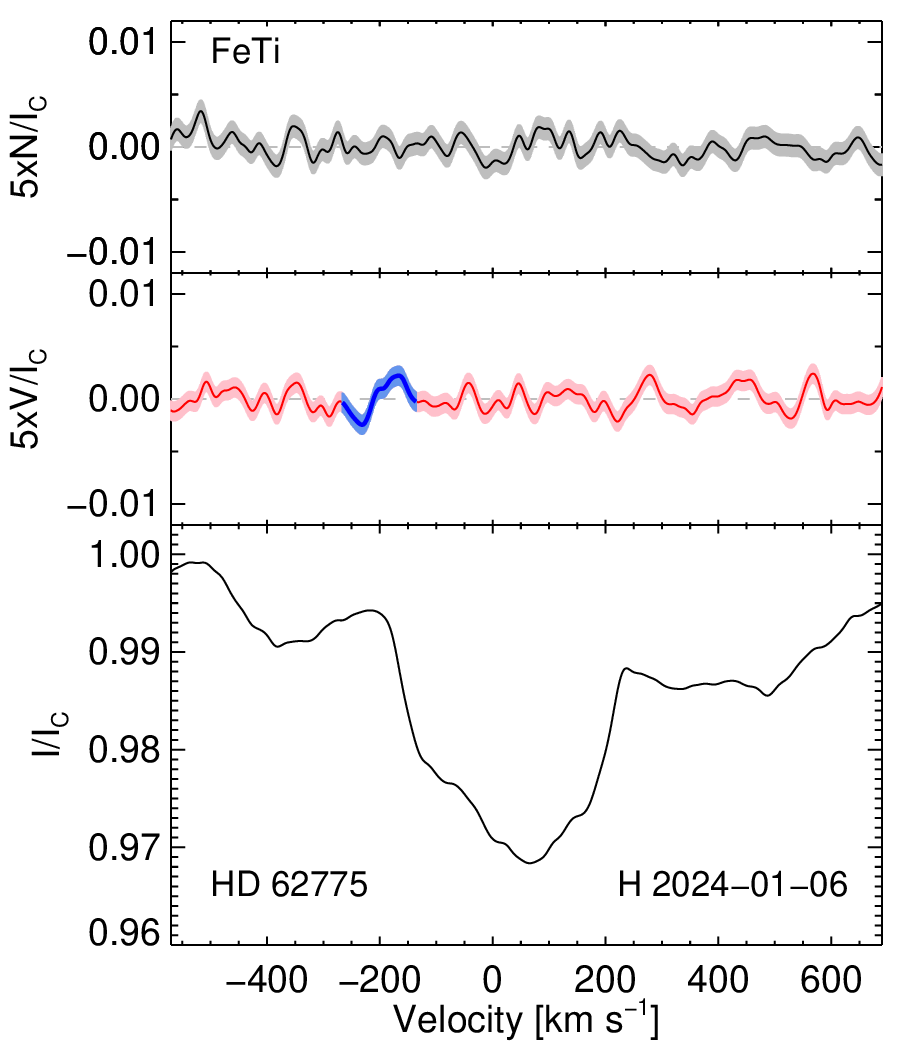}
    \includegraphics[width=0.245\textwidth]{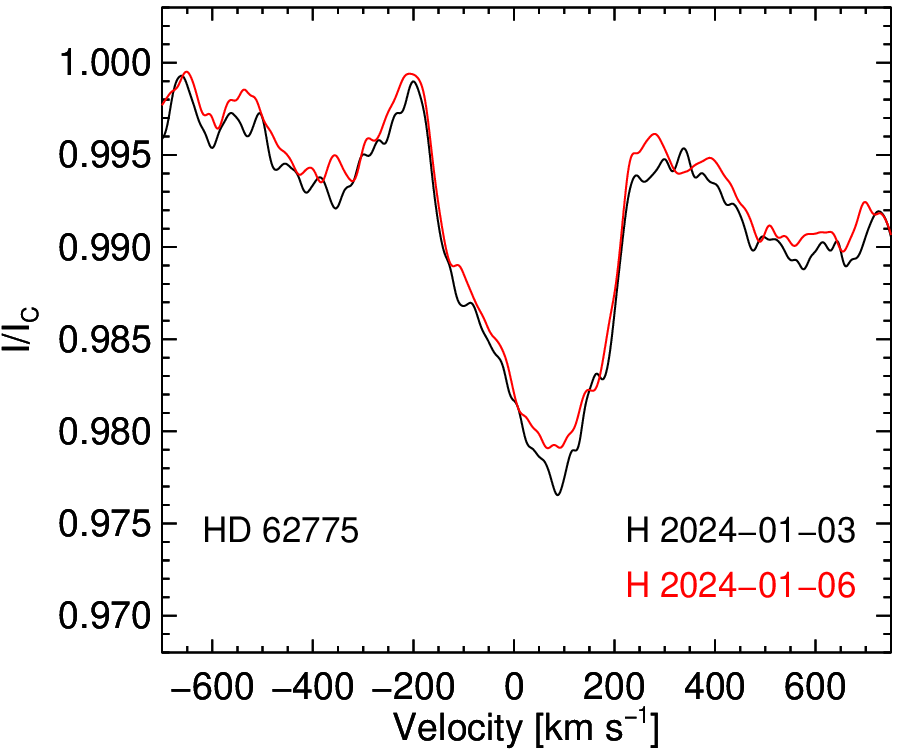}
    \caption{
As Figure~\ref{fig:IVN61954}, but for HD\,62775. On the right side we show 
overplotted LSD Stokes $I$ profiles produced for both observing epochs using a line mask with the cleanest \ion{Fe}{ii} lines.}
    \label{fig:IVNHD62775}
\end{figure*}

This target with spectral type A2\,III/IV is listed in the catalogue
of \citet{Rain2021} as a BSS in NGC\,2447 at a membership probability of 0.9. It has not been well
studied in the past, with only 12 entries in the SIMBAD database.
We observed this target on two different nights, on January 3 and January 6 2024. The results of the LSD analysis are
presented in Table~\ref{tab:obsall}. The LSD Stokes~$I$, $V$, and diagnostic null $N$ profiles obtained on January 3 and 6 2024 are presented in Fig.~\ref{fig:IVNHD62775} and indicate that this target is probably a triple system.
In the same figure on the right side 
we show the overplotted LSD Stokes $I$ profiles produced for both observing epochs using a line mask with the cleanest \ion{Fe}{ii} lines.
We do not detect significant spectral changes in either the radial velocities or the line shapes, apart from a very small
change in the depth of the central component. Clearly, more observations are necessary to firmly confirm the presence of companions.

We achieved on the first observing night a marginal detection of the mean longitudinal magnetic field using
Fe lines with $\left< B_{\rm z} \right>=631\pm110$\,G and ${\rm FAP}=6.4\times10^{-5}$ and a definite detection
with ${\rm FAP}=1.1\times10^{-7}$ using Fe and Ti lines for the second observation obtained at a higher $S/N$.
We were not able to estimate the strength of the mean longitudinal
magnetic field because it is not obvious whether the detected Zeeman signature indeed belongs to the central component and is shifted
from the line profile centre due to the presence of a surface FeTi spot.

\subsection{Trumpler\,9}
  \label{sect:mfield.trumpler9}

\paragraph*{HD\,65032 (=CD$-$25\,5284):}

\begin{figure}
    \centering
    \includegraphics[width=0.237\textwidth]{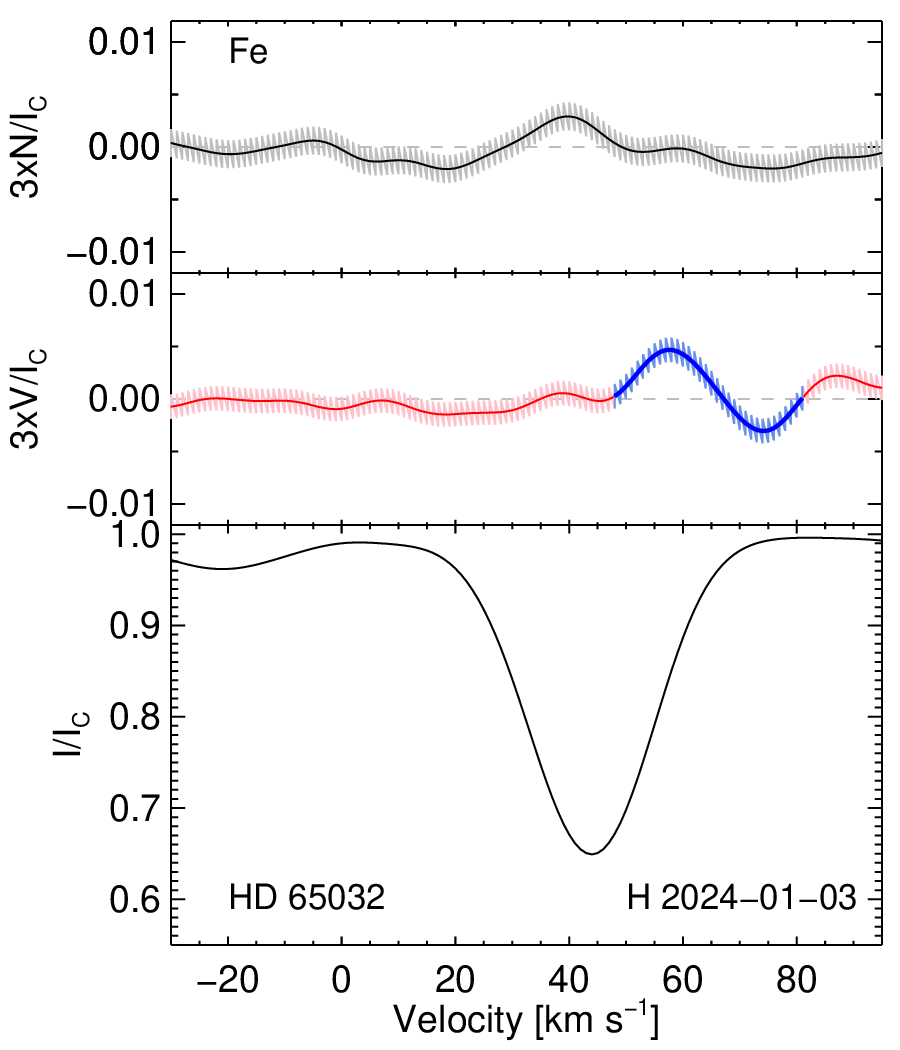}
    \includegraphics[width=0.237\textwidth]{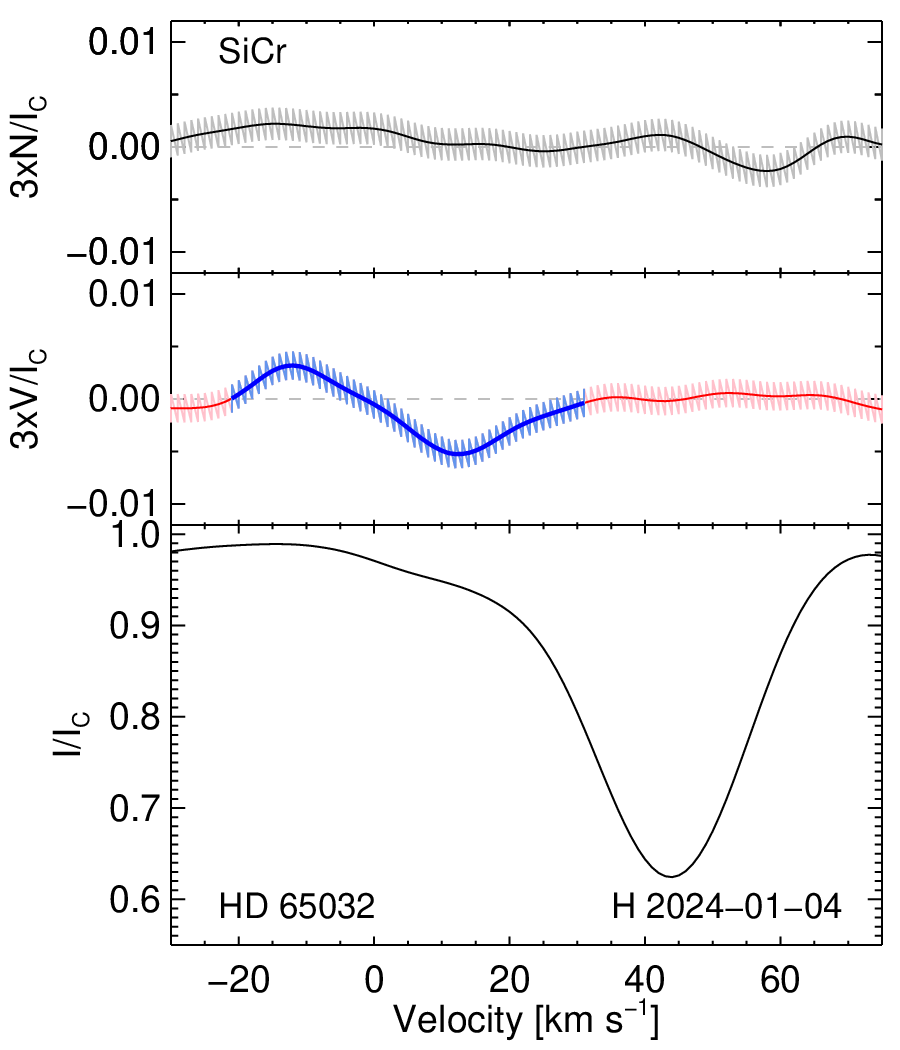}
    \caption{As Figure~\ref{fig:IVN61954}, but for HD\,65032.
}
    \label{fig:IVNHD65032}
\end{figure}

With only ten entries in the SIMBAD database, this target with spectral type A2/3\,III has been poorly studied in the past.
It is listed as a BSS in the OC Trumpler\,9 with a probability of 0.9 by \citet{Rain2021}.
However, given the Gaia $G$ magnitude of 8.249 and the colour $BP-RP$ of 0.497 \citep{Gaia2022}, HD\,65032 is clearly not a BSS but a YSS.
 This result is subject to verification because the impact of the presence of a magnetic field and 
the extent to which this influences the Gaia colours has not been investigated yet.
Two HARPS\-pol observations of this star have been acquired on January~3 and 4 2024. As is shown in Fig.~\ref{fig:IVNHD65032},
the observation from January 4 indicates the presence of a companion.
Our LSD analysis shows the marginal presence of a mean longitudinal magnetic field in the first observation with
${\rm FAP}=0.5\times10^{-5}$ using a line mask containing Fe lines. The detected Zeeman signature
is shifted from the line centre to the red wing and it is not clear whether this shift could be caused by the presence of a chemical spot with
Fe overabundance.
A definite detection of a Zeeman signature has been achieved
for the second observing night using a mask with Si and Cr lines with ${\rm FAP}=7.8\times10^{-8}$.
In this observing epoch  we see a small change in radial velocity of the order of 1\,km\,s$^{-1}$ suggesting a physical bound
  between the components. The detected Zeeman signature is shifted to the blue wing.
Obviously, more observations are necessary to be able to produce surface element maps
and to measure the magnetic field using spectral lines of inhomogeneously distributed elements separately.
Also a possible contamination by foreground or background cluster members should be investigated.
The results of our LSD analysis are presented in Table~\ref{tab:obsall} and in Fig.~\ref{fig:IVNHD65032}.

\subsection{NGC\,3114}

The cluster NGC\,3114 is located in a crowded field projected onto the Carina
complex and is reported to be rich in peculiar Ap and Bp stars \citep{Levato1975}.
On the other hand,
such a location of  NGC\,3114 complicates the separation of cluster members from field stars.
We have selected HD\,87222 and HD\,87266 for spectropolarimetric observations because 
\citet{Ahumada1995} identified both stars as blue straggler candidates and because \citet{Gonzalez2001}
reported that they are also the bluest in the turn-off zone.

\paragraph*{HD\,87222 (=ALS\,18018):}

\begin{figure}
    \centering
    \includegraphics[width=0.237\textwidth]{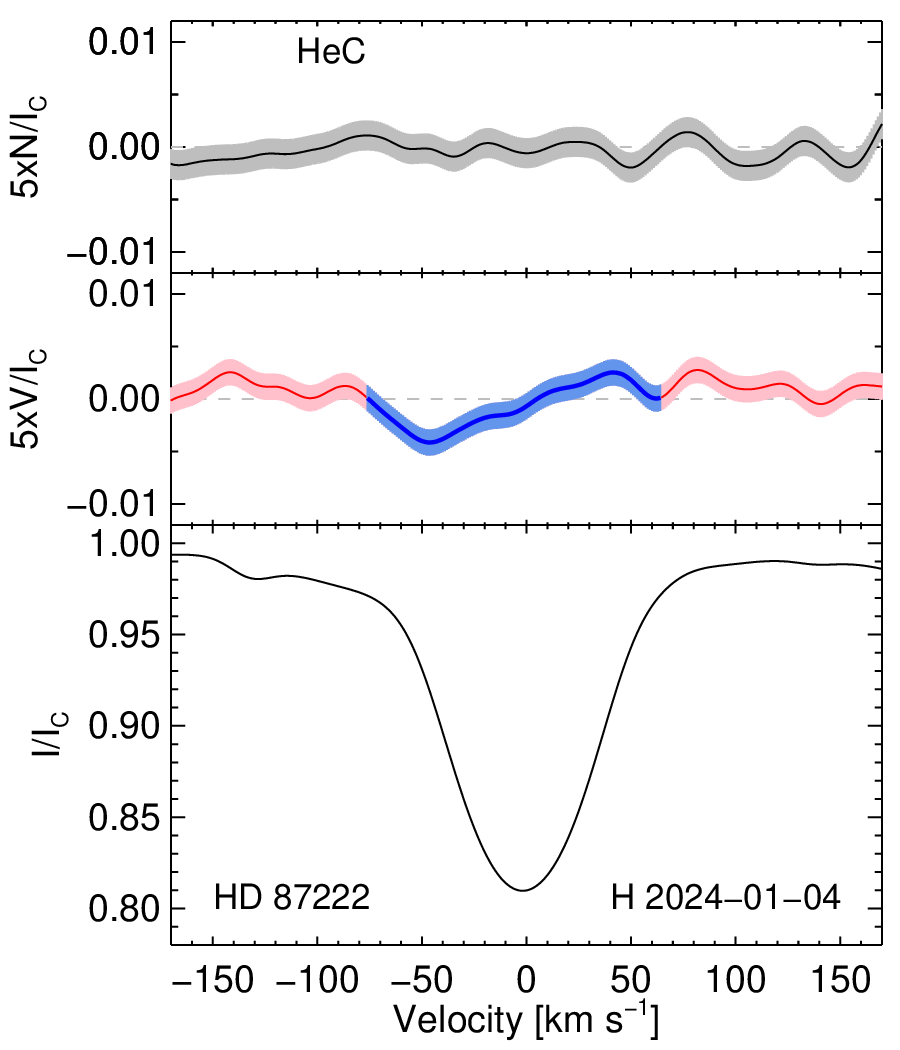}
    \includegraphics[width=0.237\textwidth]{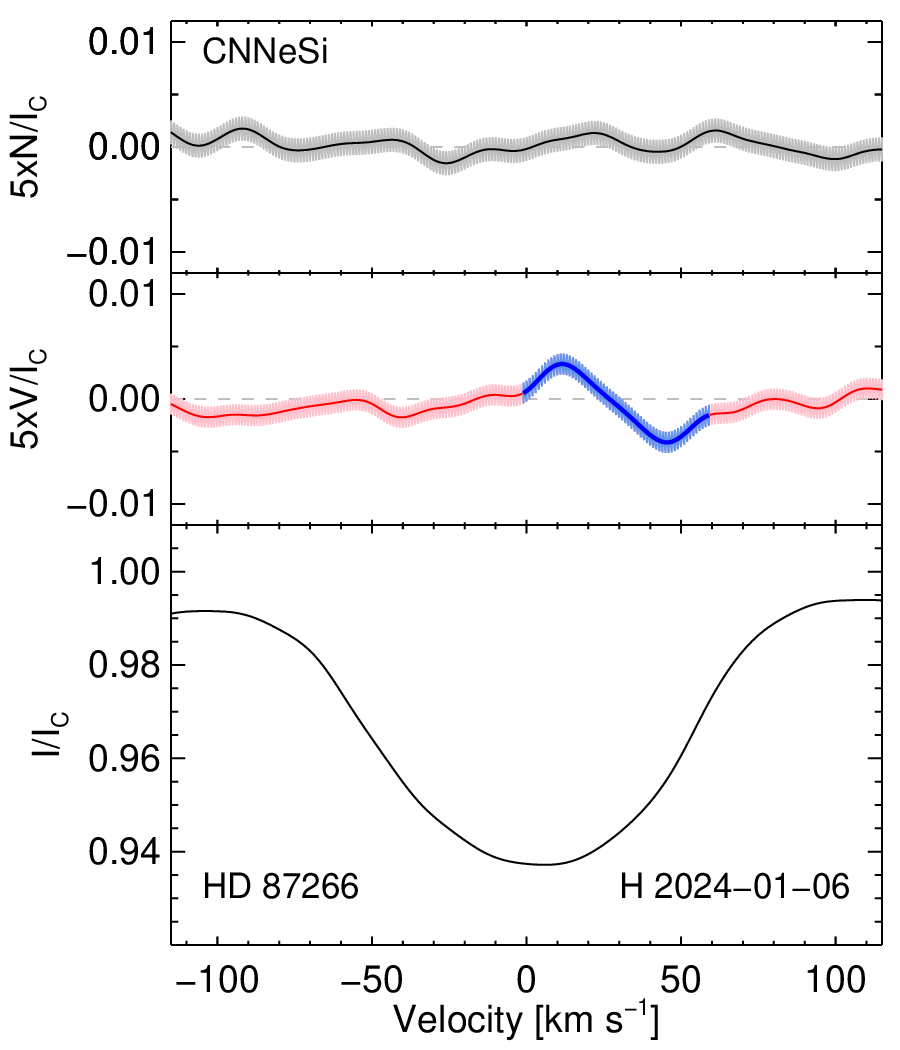}
    \caption{
{\it Left:} As Figure~\ref{fig:IVN61954}, but for HD\,87222.
{\it Right:} As Figure~\ref{fig:IVN61954}, but for HD\,87266.
}
    \label{fig:IVNHD87222}
\end{figure}

A membership probability of 99\% was reported for HD\,87222 by \citet{Gonzalez2001}, who also determined a spectral type
  B3\,V by comparison with spectra of standard stars obtained with the same instrument. This spectral classification is
  in good agreement with the classification B3\,IV reported by \citet{Aidelman2015}, who utilised the spectrophotometric
  Barbier-Chalonge-Divan (BCD) method
\citep{Barbier1941,Chalonge1952} as a tool to evaluate accurate physical parameters for cluster members.
HD\,87222 is listed in the catalogue of peculiar stars by \citet{Renson2009}. The authors assumed that
it is chemically peculiar based on the work of \citet{Levato1975}, but information about the presence of a magnetic field is still missing.
The comparison of our high-resolution
  HARPS\-pol Stokes~$I$ spectra obtained for HD\,87222 and HD\,87266
  reveals strong similarity, indicating only slightly different atmospheric parameters.
This is in agreement with the results reported for both stars by \citet{Aidelman2015}.
Our spectral classification of HD\,87222, following the criteria discussed by \citet{Negueruela2024},
suggests that the spectral type is close to B3, similar to the spectral type listed in the SIMBAD database.
To assess the chemically peculiar nature of HD\,87222,
we compared our HARPS\-pol spectrum with spectra of a few  magnetic He-rich stars and non-magnetic stars of similar spectral
type and projected rotational velocity, acquired in the framework of the ESO program 191.D-0255.
We notice that all spectral lines identifiable in the B2\,IV-V primary of
the SB2 system HD\,136504 (=$\epsilon$~Lup), detected as magnetic by \citet{Hubrig2009},
appear at a similar strength in the spectra of HD\,87222 and HD\,87266.
However, a comparison with the non-magnetic stars also does not show any clear evidence
of differences in the spectral appearance.
According to \citet{Ghazaryan2019}, the only element showing solar or anomalously strong abundance
in He-rich stars is He, whereas other elements can either be underabundant or overabundant.
However, He is usually inhomogeneously distributed over the stellar surface with a stronger concentration
at the magnetic poles (e.g.\ \citealt{Hubrig2017b}).
As HD\,87222 has been observed only once, we are not able to assess its spectral variability.
Using the line mask with He and C lines for the LSD analysis of HD\,87222, we achieve a marginal detection of a mean
longitudinal magnetic field  $\left< B_{\rm z} \right>=-133\pm7$\,G, with an FAP value of $8.8\times10^{-4}$.
The results of our analysis are presented in Table~\ref{tab:obsall} and in Fig.~\ref{fig:IVNHD87222} on the left side.

\paragraph*{HD\,87266 (=CD$-$59\,2755):}

A spectral classification B2.5\,V for this target was reported by \citet{Gonzalez2001}. Using the BCD method, \citet{Aidelman2015}
  reported a spectral type of B2\,III. Our spectral classification following the criteria discussed by \citet{Negueruela2024} indicates
  that the spectral type is closer to B2.5.
HD\,87266 is mentioned as a BSS in the catalogue of
\citet{Ahumada1995} and, similar to HD\,87222, is listed in the \citet{Renson2009} catalogue as a peculiar star,
again without mentioning a magnetic field.
Similar to HD\,87222, HD\,87266 has been observed only once. Thus, we are not able to assess its spectral  variability.
HD\,87266 belongs to NGC\,3114 with a probability of 0.9 \citep{Cantat2020}.
Although previous spectropolarimetric observations
failed to detect a magnetic field in HD\,87266 (e.g.\ \citealt{Bagnulo2006}), our HARPS\-pol observation
shows a definite longitudinal magnetic field detection:  $\left< B_{\rm z} \right>=110\pm23$\,G, with
an FAP value of $1.0\times10^{-5}$, if we employ a line mask containing C, N, Ne, and Si lines.
The shifted location of the observed
Zeeman signature in the LSD Stokes~$V$ spectrum probably indicates that there are surface chemical spots
or that HD\,87266 is a close binary with a magnetic component.
The results of our analysis are presented in Table~\ref{tab:obsall} and in
Fig.~\ref{fig:IVNHD87222} on the right side.

\subsection{IC\,2944}

\paragraph*{HD\,101545\,A (=ALS\,2448):}

\begin{figure}
    \centering
    \includegraphics[width=0.237\textwidth]{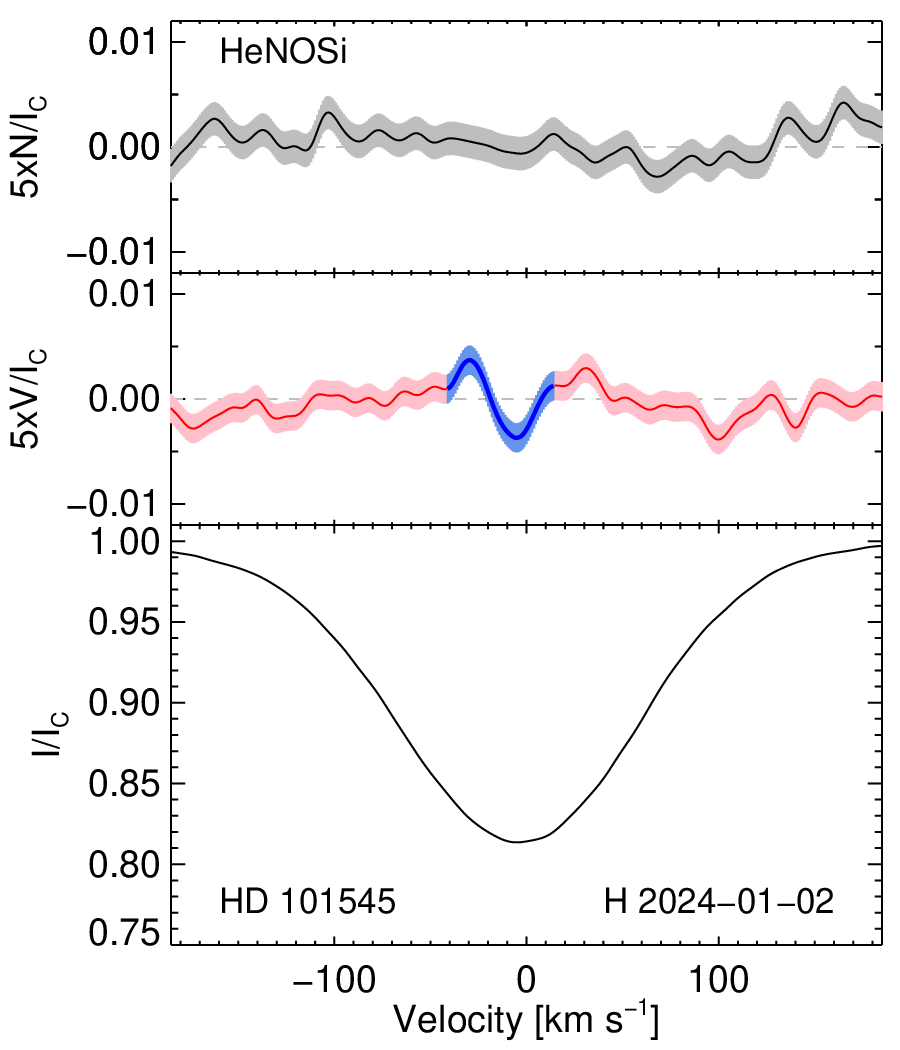}
    \includegraphics[width=0.237\textwidth]{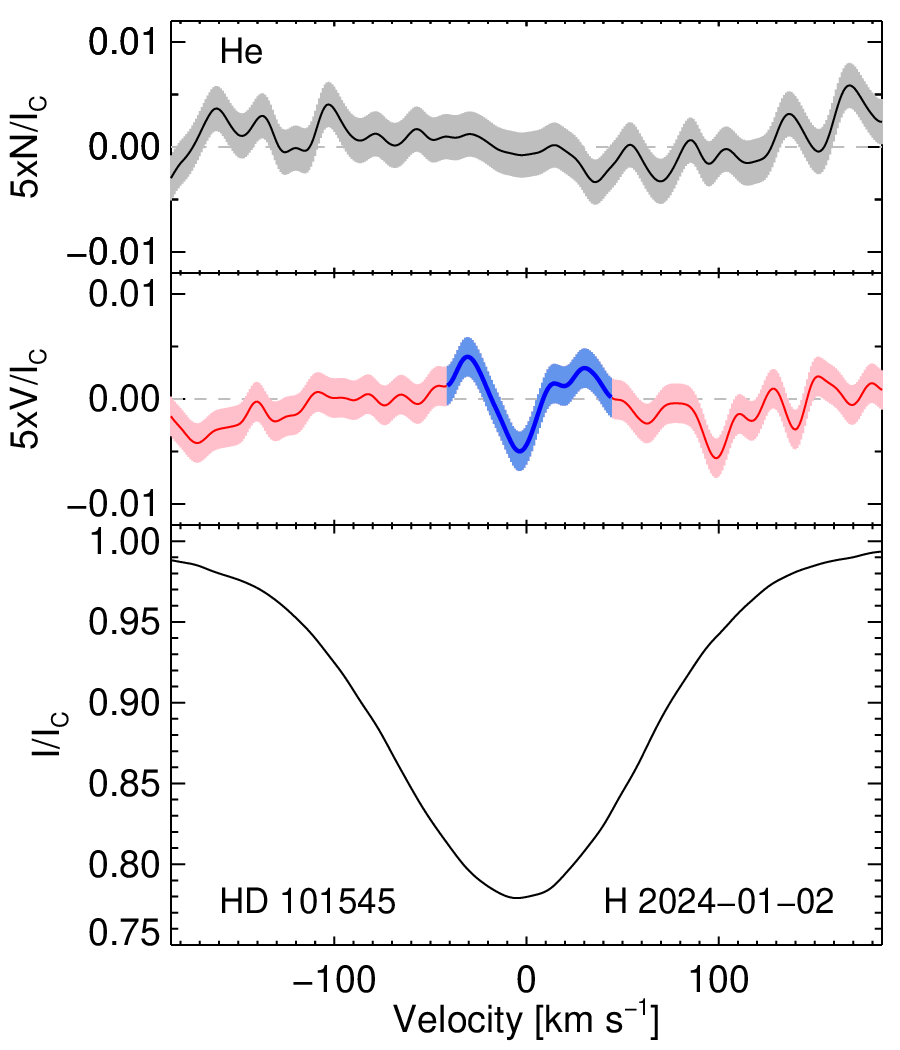}
    \caption{As Figure~\ref{fig:IVN61954}, but for HD\,101545\,A.
}
    \label{fig:IVNHD101545}
\end{figure}

This is the primary in the O9.5Ib$+$B0.5Iab visual binary with a component separation of 2.6\arcsec{} \citep{Sana2014} and
has been identified as a BSS in IC\,2944 in the work of \citet{Ahumada2007}.
However, as was mentioned above, this cluster was not included
in the catalogue of \citet{Rain2021}, probably because of limitations set by photometric calibration errors for bright targets.
According to \citet{Baumgardt2000}, the membership probability for HD\,101545 is 0.83.
The study of \citet{Sana2011} indicates that the best spectral type estimate for this system yielded
O9.5\,III/I and O9.7 III/I for the A and B components, but given the
stars' brightness, the authors finally considered both stars to be giants and not
supergiants.
No magnetic field measurements for this system have been carried out
in the past. We observed HD\,101545\,A with HARPS\-pol only once on January~2 2024 and report here the definite detection
of a mean longitudinal magnetic field.
In our analysis we employed two different masks, one mask containing He, N, O, and Si lines and one mask containing
exclusively He lines. For the mask with the He, N, O, and Si lines we obtain a marginal detection
$\left< B_{\rm z} \right>=69\pm17$\,G with ${\rm FAP}=2.1\times10^{-5}$.
For the mask with the He lines, we obtain a definite detection of the mean longitudinal field,
$\left< B_{\rm z} \right>=82\pm16$\,G, with ${\rm FAP}=4.3\times10^{-6}$.
The results of our analysis are presented in Table~\ref{tab:obsall} and in
Fig.~\ref{fig:IVNHD101545}.

\section{Discussion}
\label{sect:disc}

In this work, we present the first observational evidence that BSSs and YSSs possess magnetic fields of the order of a hundred Gauss.
By analysing the polarized spectra of the sample targets using
the LSD technique, we extracted Zeeman signatures that revealed definite detections of the presence of a magnetic field
in four BSSs and three YSSs. For the BSS HD\,87222, only a marginal detection has been achieved.
Importantly, the detection of the presence of
a magnetic field in YSSs with a field strength comparable to that observed in BSSs indicates
that magnetic fields created during the binary interaction process probably remain
stable on evolutionary timescales. 

For all targets apart from HD\,87266, the magnetic field measurements have been carried out for the first time.
The measured mean longitudinal magnetic field strength for most targets is of the order of a hundred to a few hundred Gauss.
Only for the massive target HD\,101545 is it below 100\,G. However, our targets have been observed only once or twice.
As the longitudinal magnetic field is strongly dependent on the
viewing angle between the field orientation and the observer and is modulated as the star rotates, spectropolarimetric monitoring
over the rotation periods of our targets is necessary to obtain
trustworthy statistics on their magnetic field geometries and the distribution of the field strengths.
Importantly, it is the first time that, using high-resolution spectropolarimetric observations,
  a definite detection of a magnetic field has been achieved in a Be-shell star, HD\,61954.
  The only other magnetic Be-shell star with a definite detection, HD\,56014, hosting a longitudinal
magnetic field  $\left< B_{\rm z} \right>=-146\pm32$\,G, was reported by
\citet{Hubrig2007}. In their study, the authors used low-resolution observations with the ESO multi-mode instrument FORS1
installed at the 8\,m Kueyen telescope.

Apart from HD\,101545, which is a known visual binary, we have discovered that our targets,
HD\,62329 and HD\,65032, appear to be members of binary systems and that HD\,62775 is possibly a member in a triple system.
Because the amplitudes of the Zeeman signatures are lower in multiple systems in comparison to the size of
these features in single stars (e.g.\ \citealt{Hubrig2023}), their magnetic fields can even be stronger.
The multiplicity of the three targets mentioned above has not previously been mentioned in the literature. 
Unfortunately, only single observations have been obtained for the remaining targets: HD\,62000, which has a very unusual
spectral appearance exhibiting shell-like line profiles, HD\,87222, and HD\,87266,
all of which appear as single stars in our spectra.
It is not clear whether they are merger products or we are unable to detect their companions due to the small number of observations.
The presence of a companion can certainly remain
undetected for some targets when only single or very few observations are available or when spectropolarimetric observations
are obtained with low $S/N$.
Still, our detection of a few binary and multiple systems supports the theoretical scenario that magnetic
fields may be generated by strong binary interactions.
These results are also in agreement with the results of \citet{Hubrig2023}, who demonstrated that magnetic fields are
frequently observed in binary and multiple systems with O- and B-type components. 

\begin{figure}
    \centering
    \includegraphics[width=0.237\textwidth]{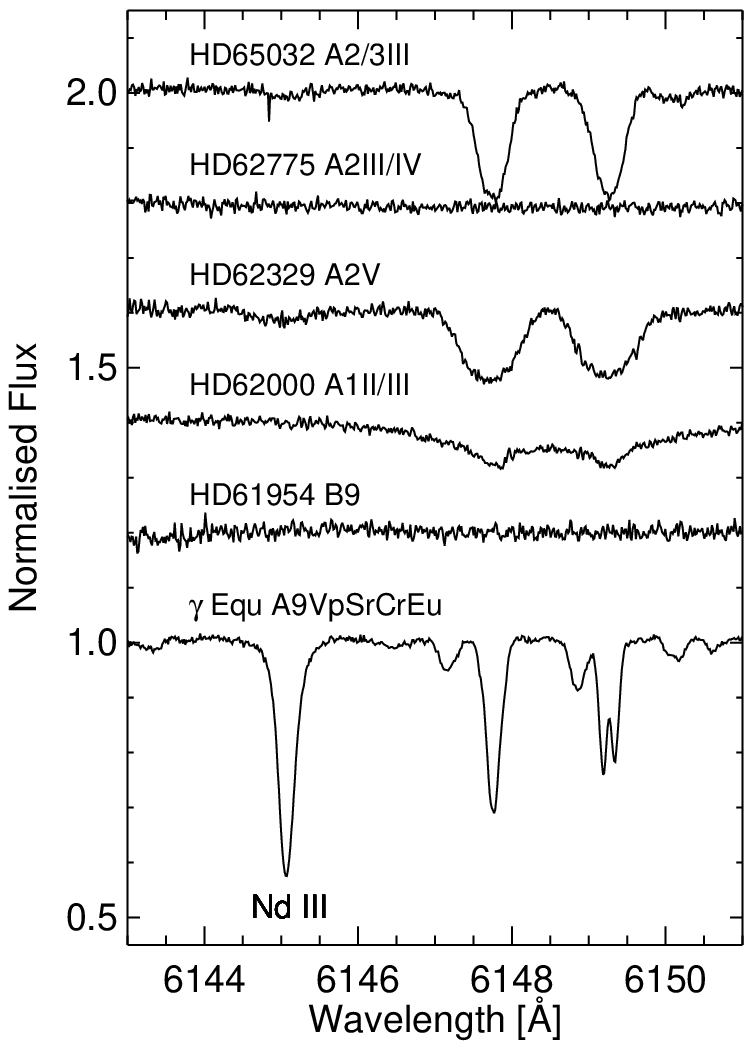}
    \includegraphics[width=0.237\textwidth]{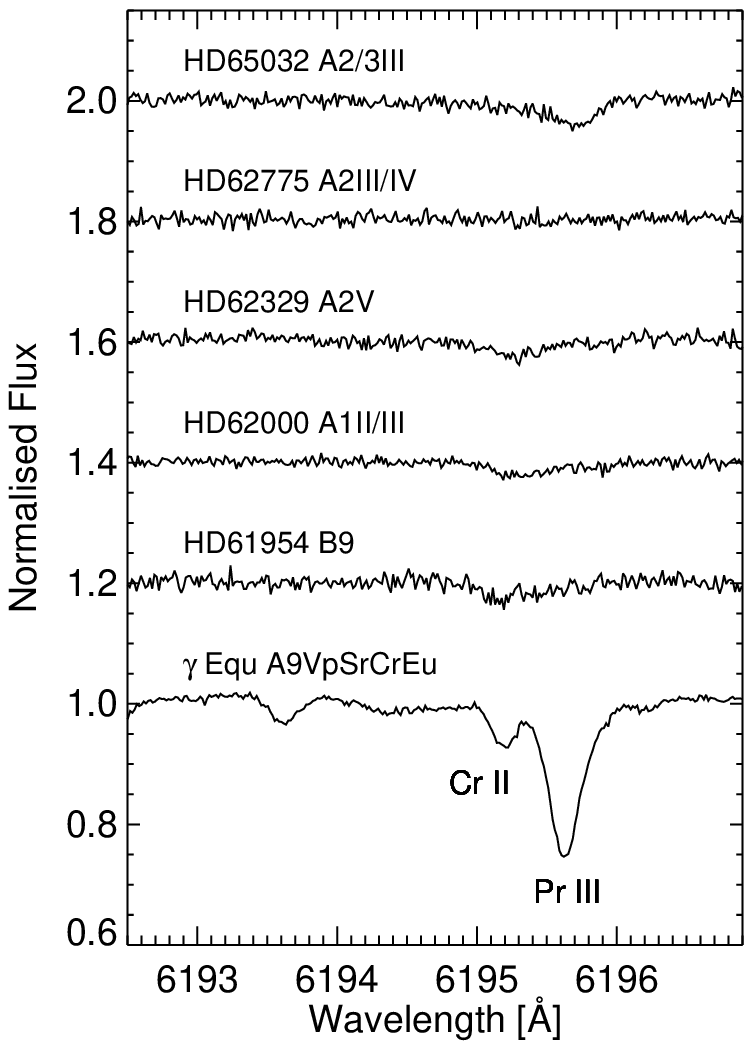}
    \caption{
      Spectra of our late-B and A-type targets in the spectral regions containing the spectral lines
\ion{Nd}{iii}~6145.1 and \ion{Pr}{iii}~6195.6.
      For comparison, we show in the bottom the spectrum of the typical Ap star $\gamma$~Equ.
}
    \label{fig:rearth}
\end{figure}

Importantly, magnetic A and late-B type stars, generally called Ap and Bp stars, are frequently severely enhanced in most of the rare earth
elements (REEs).
Rare earth elements are usually concentrated in surface spots close to the magnetic field poles. 
For the majority of Ap and Bp stars, the spectral lines \ion{Nd}{iii}~6145.1 and \ion{Pr}{iii}~6195.6 appear especially prominent in their spectra.
In Fig.~\ref{fig:rearth} we present the spectra of our targets in the spectral regions containing these lines.
In the bottom of the two panels, each for a different spectral wavelength region, we display
for comparison the spectrum of the typical Ap star $\gamma$~Equ. While the very weak \ion{Nd}{iii}~6145.1 line is detectable
in the spectra of HD\,65032 and HD\,62329, the presence of the \ion{Pr}{iii}~6195.6 line is detectable only in HD\,65032.
Since the visibility of the rare earth spots is changing over the course of 
the stellar rotation, additional observations over different rotational phases are necessary to prove that the other targets
indeed do not show similar REE enhancement. It is also not clear on which timescales the development of the element overabundance due to
the atomic radiative diffusion process involving the interplay of gravity and radiation pressure takes place.

Ap and late Bp stars are also very rarely members of close binary and
multiple systems, but frequently members of wide binaries (e.g.\ \citealt{Mathys2017}). Among a sample of 113 Ap stars studied by
\citet{Carrier2002}, at least 34 binaries were found. After correcting for detection biases, the authors estimated an Ap binary fraction
of 43\%. \citet{Mathys2017} confirmed 21 of the 84 Ap stars to be
binaries and reported that the shortest periods among them were $P_{\rm orb}= 3.37$\,d for HD\,142070 and $P_{\rm orb}= 8.03$\,d
for HD\,65339.
The $P_{\rm orb}$ of the remaining binaries were all $>$18\,d, and half had $P_{\rm orb}>$1000\,d. This study is raising the
important question of what happened to the short orbital period Ap stars, and whether it is possible that BSSs,
which are usually considered as coalesced stars or 
rejuvenated companions in binary systems due to mass transfer, present the missing short period binaries among the Ap stars.
It is of interest  that nearly all main-sequence late B-type stars in close binaries with $P_{\rm orb} \le 20$\,d show HgMn peculiarities
and possess only weak magnetic fields (e.g.\ \citealt{Hubrig1995, Hubrig2012}),
whereas A type stars in close binaries frequently show
a weak metal overabundance and are correspondingly classified as Am stars (e.g.\ \citealt{Kitamura1978}). 

Although the targets in our sample belong to OCs of very different ages
and metallicities (see Table~\ref{tab:clusters}), we do not detect any relationship  between the presence or strength of the detected
magnetic field and the cluster characteristics.
Interestingly, magnetic fields have recently also been detected in massive stars in the low-metallicity environment
of the Magellanic Clouds \citep{Hubrig2024}, indicating that the impact
of the lower-metallicity environment on the occurrence and strength of stellar magnetic fields in massive stars is
low. On the other hand, since the number of the studied BSSs and YSSs is small, 
spectropolarimetric observations of a representative sample of BSSs
are necessary to provide more reliable results about possible correlations with cluster parameters.

Considering the catalogue of \citet{Rain2021}, only 13 BSSs with $m_{\rm V}$ < 10  are observable from
La~Silla, i.e.\ with declinations below $+20^{\circ}$, and have ages less than 700\,Myr.
Among the YSSs, only six targets fulfil these criteria.
Follow-up studies of the presence of a magnetic field in a larger sample of BSSs and YSSs will be worthwhile
to provide the crucial information necessary to test predictions of existing theories and
place more stringent constraints on the origin of magnetic fields in intermediate-mass and massive stars. 

\begin{acknowledgements}

We appreciate the constructive comments by the referee and
thank Rahul Jayaraman for helpful discussions.
This work is based on observations made with ESO telescopes at the
La Silla Paranal Observatory under programmes ID 0112.D-0091(A) and
ID 191.D-0255 publicly available via the ESO Archive.

\end{acknowledgements}

%
   \bibliographystyle{aa} 
   \bibliography{aa56166-25} 
%

\end{document}